\documentclass[12pt]{article}
\usepackage{natbib}
\usepackage{amsmath,amssymb}
\usepackage[a4paper,tmargin=3cm,bmargin=2cm,rmargin=2cm,lmargin=2cm]{geometry}
\usepackage{graphicx,color}
\usepackage{multirow}
\usepackage[normalem]{ulem}

\DeclareMathOperator{\Pc}{F}

\newcommand{\N}[1]{#1}
\newcommand{\D}[1]{}

\begin{document}
\title{Thermal and compositional stratification of the inner core}
\author{St\'ephane Labrosse\\
{\small Laboratoire de G\'eologie de Lyon, ENS de
Lyon, Universit\'e Lyon-1, Universit\'e de Lyon}\\
{\small 46 all\'ee d'Italie, 69007 Lyon, France}\\
{\small Tel: +33 4 72 72 85 15. Fax: +33 4 72 72 86 77.}\\
{\small stephane.labrosse@ens-lyon.fr}\\
{\small Institut Universitaire de France. }}
\maketitle
\begin{abstract}
  The improvements on the knowledge of the seismic structure of the
  inner core and the complexities thereby revealed ask for a dynamical
  origin. Sub-solidus convection was one of the early suggestions to
  explain the seismic anisotropy but requires an unstable density
  gradient either from thermal or compositional origin, or
  both. Temperature and composition profiles in the inner core are
  computed using a unidimensional model of core evolution including
  diffusion in the inner core and fractional crystallisation at the
  the inner core boundary (ICB). The thermal conductivity of the core
  has been recently revised upwardly and, moreover, found increasing
  with depth. Values of the heat flow across the core mantle boundary
  (CMB) sufficient to maintain convection in the whole outer core are
  not sufficient to make the temperature in the inner core
  super-isentropic and therefore prone to thermal instability. An
  unreasonably high CMB heat flow is necessary to this end. The
  compositional stratification results from a competition of the
  increase of the concentration in light elements in the outer core with
  inner core growth, which makes the inner core concentration also
  increase, and of the decrease of the liquidus which makes the
  partition coefficient decrease as well as the concentration of light
  elements in the solid. While the latter (destabilizing) effect
  dominates at small inner core sizes, the former takes over for a
  large inner core. The turnover point is encountered for an inner
  core about half its current size in the case of S but much larger
  for the case of O. The combined thermal and compositional buoyancy
  is stabilizing and solid-state convection in the inner core appears
  unlikely, unless an early double-diffusive instability can set in.
\end{abstract}


\section{Introduction}
\label{sec:introduction}
Since the discovery of the inner core anisotropy
\citep{Morelli_etal1986,Poupinet1983,Woodhouse_etal1986}, many
different mechanisms have been proposed to explain this observation
\citep{Deguen2012}. It is not clear at present if any of these models
is in fact able to quantitatively explain the observations and it is
necessary to test systematically all the scenarios. This paper deals
with one of the first proposed scenario: convection in the inner core
\citep{Buffett2009,Cottaar_Buffett2012,Deguen_Cardin2011,Deguen_etal2013,Jeanloz_Wenk88,Mizzon_Monnereau2013,Weber_Machetel92}.

Convection in the solid inner core is possible, like in the solid
mantle, provided a sufficient source of buoyancy is available. For
thermal convection to occur, the buoyancy source must come from a
combination of radiogenic heating
\citep{Jeanloz_Wenk88,Weber_Machetel92}, secular cooling
\citep{Buffett2009,Deguen_Cardin2011,Cottaar_Buffett2012,Deguen_etal2013,Mizzon_Monnereau2013}
or even Joule heating \citep{Takehiro2011}. The
amount of potassium in the core is likely very limited
\citep{Hirose_etal2013} and will not be considered further. 
Joule heating in the inner core \citep{Takehiro2011} depends on the strength and pattern of the magnetic
field at the bottom of the outer core and will also be omitted here.
Secular
cooling can provide enough buoyancy to drive thermal convection in the
inner core if the cooling is fast enough compared to the time required
to cool the inner core by diffusion.
This question was
investigated in great details in a few recent papers
\citep{Yukutake1998,Buffett2009,Deguen_Cardin2011,Deguen_etal2013}. In particular,
\citet{Deguen_Cardin2011} proposed an approximate criterion for the
possibility of thermal instability involving the age of the inner core
and the thermal conductivity of the inner core. Recent results on the
thermal conductivity of the core
\citep{deKoker_etal2012,Pozzo_etal2012,Gomi_etal2013,Pozzo_etal2014} favor a value
much larger than previously thought which makes the case for inner
core thermal convection harder to defend. This will be discussed in
section~\ref{sec:cond-therm-conv}. 

Compositional convection is also possible if the metal that
crystallizes at the inner core boundary (ICB) gets depleted in light
elements as the inner core grows. The
concentration in light element $X$ in the solid, $C_X^s$, is related
to that of the liquid $C_X^l$ by 
\begin{equation}
  \label{eq:partition}
  C_X^s=P_X^{sl} C_X^l,
\end{equation}
$P^{sl}$ being the partition coefficient, \N{generally lower than
  1}. As discussed by \citet{Deguen_Cardin2011}, $C_X^s$ can vary
because of the variation of $P_X^{sl}$ and $C_X^l$. Assuming that the
outer core is compositionally well mixed, $C_X^l$ increases with the
inner core growth due to the expulsion from the inner core with
$P_X^{sl}<1$. This effect tends to create a stably stratified inner
core and must be compensated by a decrease of $P_X^{sl}$ for
compositional convection to occur. \citet{Gubbins_etal2013} proposed
that the decrease of the liquidus temperature with inner core growth
is able to provide such a variation. This effect will be discussed in
section~\ref{sec:comp-strat}. The combined thermal and compositional
buoyancy will then be discussed in section~\ref{sec:comb-buoy-prof}.

Compared to the previous work cited above, this paper differs in
several ways. I do not attempt to solve the full convection problem as
done by \citet{Deguen_Cardin2011} and \citet{Deguen_etal2013} because
I merely want to study the conditions under which the basic
stratification in a diffusion regime can become unstable, conditions
that are found hard to meet with the large thermal conductivity
implied by the recent studies. On the other hand, I solve the full
thermal diffusion problem including the moving inner core boundary,
coupled to the outer core evolution, which was not done by the
previous workers on the topic\N{, except \citet{Yukutake1998} who did
  not consider compositional effects}. The compositional evolution
follows from the thermodynamics relations of \citet{Alfe_etal02a} and
\citet{Gubbins_etal2013} but are treated in a more self-consistent way
than the latter study, as discussed below.

\section{Model for the evolution of the \N{inner} core}
\label{sec:numerical-model}
\begin{table}[htbp]
  \centering
  \begin{tabular}{llll}
    \hline
    \textbf{Parameter} & \textbf{Symbol} & \multicolumn{2}{l}{\textbf{Value}} \\
    Boltzmann constant &$k_B$ & \multicolumn{2}{l}{$8.617\
      10^{-5}\mathrm{eV\ atom^{-1}}$}\\
    Core radius$^a$ & $r_{OC}$ & \multicolumn{2}{l}{$3480\ \mathrm{km}$}\\
   Present inner core radius$^a$ & $r_{ICf}$ & \multicolumn{2}{l}{$1221\
      \mathrm{km}$}\\
    Density length scale$^b$ & $L_\rho$ & \multicolumn{2}{l}{$7680 \mathrm{km}$}\\
    Thermal expansion coefficient & $\alpha$ &
    \multicolumn{2}{l}{$10^{-5}\mathrm{K^{-1}}$}\\
    Heat capacity$^c$ &$C_P$& \multicolumn{2}{l}{$750\mathrm{J
        K^{-1}kg^{-1}}$}\\
    Present ICB temperature, M\&G model$^d$ & $T_L(r_{ICf})$ & \multicolumn{2}{l}{$5500\mathrm{K}$}\\
    Present ICB temperature, PREM model$^d$ & $T_L(r_{ICf})$ & \multicolumn{2}{l}{$5700\mathrm{K}$}\\
    Present CMB isentropic heat flow, M\&G model & $Q_s(0)$ & \multicolumn{2}{l}{$13.3\mathrm{TW}$}\\
    Present CMB isentropic heat flow, PREM model & $Q_s(0)$ & \multicolumn{2}{l}{$13.8\mathrm{TW}$}\\
    Compositional dependence of the liquidus$^e$ &
    $\left(\frac{\partial T_L}{\partial\xi_O}\right)_{P}$ &
    \multicolumn{2}{l}{$-21\ 10^3\mathrm{K}$}\\
    Pressure dependence of the liquidus temperature$^f$ &
    $\left(\frac{\partial T_L}{\partial P}\right)_{\xi_O}$ &
    \multicolumn{2}{l}{$9\mathrm{K\ GPa^{-1}}$}\\[0.5cm]
    Thermal conductivity at the center$^g$ & $k_0$ &
    \multicolumn{2}{l}{$163\mathrm{W\ m^{-1}K^{-1}}$}\\
    Radial dependence of conductivity$^g$ & $A_k$ & \multicolumn{2}{l}{$2.39$}\\[0.5cm]
    \textbf{Parameter} & \textbf{Symbol} & \textbf{Value for O} & \textbf{Value for S}\\
    Difference in chemical potential$^d$ ($\mathrm{eV\ atom^{-1}}$) &
    $\mu_0^l-\mu_0^s$ & $-2.6$ & $-0.25$\\
    Linear correction, solid$^d$ ($\mathrm{eV\ atom^{-1}}$) & $\lambda_X^s$ & 0 & 5.9\\
    Linear correction, liquid$^d$ ($\mathrm{eV\ atom^{-1}}$) & $\lambda_X^l$ & 3.25 & 6.15\\
    Chemical expansion coefficient$^h$ & $\beta_X$ & -1.3 & -0.67\\
    Starting mass fraction in the liquid, M\&G model$^d$ & $\xi_X^l$ & $4.06\%$ & $5.18\%$\\
    Starting mass fraction in the solid, M\&G model & $\xi_X^s$ & $0.05\%$ & $3.63\%$\\
    Starting mass fraction in the liquid, PREM model$^d$ & $\xi_X^l$ & $2.42\%$ & $6.30\%$\\
    Starting mass fraction in the solid, PREM model & $\xi_X^s$ & $0.02\%$ & $4.75\%$\\
    \hline
  \end{tabular}
  \caption{Parameter values. 
    $^a$ from PREM\protect\citep{PREM}.
    $^b$ from a fit to PREM. 
    $^c$ from \citet{Gubbins_etal03}.
    $^d$ from \citet{Alfe_etal02a,Gubbins_etal2013}. Different compositions give
    different values in the parameters. The compositions are derived
    by \citet{Gubbins_etal2013} to match the density jump across the
    ICB, as found in PREM \citep{PREM} or in 
    \citet{Masters_Gubbins03} (M\&G model). Concentration are
    transformed in mass fraction, as explained in appendix. \N{The
    initial values are computed so that the final ones match those
    from \citet{Gubbins_etal2013}.}
    $^e$ \N{calculated from the molar concentration equivalent in} \citet{Alfe_etal2007}.
    $^f$ from \citet{Alfe_etal99}.
    $^g$ from \citet{Gomi_etal2013} assuming
    the most conservative value of $k_{CMB}=90\mathrm{W/m/K}$.
    $^h$ derived by \citet{Deguen_Cardin2011} from the molar
    equivalent in \citet{Alfe_etal02a}.}
  \label{tab:parameters}
\end{table}

Following \citet{Alfe_etal02a} and \citet{Gubbins_etal2013}, I assume
a ternary composition for the core with Fe, O, and S. \N{A alternative
ternary composition with Fe, O and Si will be briefly discussed in
section~\ref{sec:disc-other-poss} for completeness.} Following
\citet{Gubbins_etal2013}, two compositional models are considered, one
matching the ICB density jump of PREM \citep{PREM} (thereafter termed
PREM model) and the other one matching the ICB jump proposed by
\citet{Masters_Gubbins03} (M\&G model), which is larger. Because only
O significantly fractionate at the ICB, the larger the density jump,
the more O is needed in the core. Considering these two models allows
to investigate the implications this has on the stratification of the
inner core. 

O is highly incompatible in the inner core ($P_O^{sl}\ll 1$) while S has
a partition coefficient only slightly lower than 1, which means that
both are not very promising to create an unstable stratification in the
inner core. Indeed, the limit $P=0$ allows no solute in the inner core
and $P=1$ forbids its change in the outer core and therefore in the
inner core. In both end-member cases, no concentration stratification
is possible in the inner core and the optimum value for such a
stratification is $P=0.5$ \citep{Deguen_Cardin2011}. 

The evolution of concentrations of O and S in the outer core
from inner core growth follows from their conservation equations. These
are most readily written using their mass fraction, $\xi_X^i$, $i$
being either $s$ for solid or $l$ for liquid and $X$ any of the two
light elements considered, S or O. In the following, an omitted $X$
means that it applies to either of the two. The relation between mass
and molar fractions in the ternary system are given in
appendix~\ref{sec:mass-molar-fraction}. In terms of mass fraction, the
partition between liquid and solid is expressed by the factor $K_X^{sl}$
defined as the ratio of the mass fraction in the solid to that in the
liquid:
\begin{equation}
  \label{eq:partitionK}
  K_X^{sl}=\frac{\xi_X^s}{\xi_X^l}.
\end{equation}
The conservation of light element $X$ can simply be stated as
\begin{equation}
  \label{eq:Conservation-xi}
  \frac{d}{dt}\left(\xi^l M_{OC}\right)=-K^{sl}\xi^l \frac{dM_{IC}}{dt}
\end{equation}
\N{which expresses that the total mass of the light element in the
outer core, $\xi^l M_{OC}$, $M_{OC}$ being the outer core mass,
decreases because of the flux of solute going in the growing inner
core. For an infinitesimal duration $\delta t$, the inner
core mass increases by $\delta M_{IC}$ and incorporates a total mass
of solute equal to $\xi^s \delta M_{IC}=K^{sl}\xi^l \delta M_{IC}$.}
The total mass of the core $M_{tot}=M_{IC}+M_{OC}$ being
constant, equation~\eqref{eq:Conservation-xi} can be recast as
\begin{equation}
  \label{eq:dxidt}
  \frac{d\xi^l}{dt}=\xi^l\frac{1-K^{sl}}{M_{OC}}4\pi r_{IC}^2\rho(r_{IC})\frac{dr_{IC}}{dt},
\end{equation}
where all terms on the right-hand-side vary with time, or more
precisely with the growth of the inner core (radius $r_{IC}$, $r_{OC}$
for the outer core). Because of the very small value of the partition
coefficient for O, very little is incorporated in the inner core and I
assume $K_O^{sl}=0$ to compute the evolution of the concentration in
the outer core. The solution to equation~\eqref{eq:Conservation-xi} is
then
\begin{equation}
  \label{eq:xi-O}
  \xi_O^l=\xi_{O0}^l\frac{M_{tot}}{M_{OC}}
\end{equation}
$\xi_{O0}^l$ being the initial mass fraction of O in the core.  The
variation of $\xi_O^l$ with the inner core growth comes only from the
variation of the outer core mass and, assuming a polynomial dependence
of the density on radius of the form \citep{Gomi_etal2013}, 
\begin{equation}
  \label{eq:density}
  \rho=\rho_0\left(1-\frac{r^2}{L_\rho^2}\right),
\end{equation}
$L_\rho$ being a density length scale \citep{Labrosse_etal01}, one
gets to leading order
\begin{equation}
  \label{eq:xi-O-dev}
  \xi_O^l=\xi_{O0}^l\left[1+\frac{r_{IC}^3}{r_{OC}^3\left(1-\frac{3r_{OC}^2}{5L_\rho^2}\right)}\right].
\end{equation}
In the numerical model, a higher order expression is in fact used, as
is discussed elsewhere \citep{Labrosse2014a}. 

For the more general case where $K^{sl}$ is neither 0 nor 1 and dependent
on temperature and concentrations
\citep{Alfe_etal02a,Gubbins_etal2013}, equation~\eqref{eq:dxidt} has
to be solved numerically. Even the value of $K^{sl}$ must computed
numerically. We follow here the theory of \citet{Alfe_etal02a} and use
the same parameters (table~\ref{tab:parameters}). For this reason, the molar
partition coefficient is more convenient here. The equilibrium at the
ICB requires the chemical potential in the solid and the liquid to be
equal, that is \citep{Alfe_etal02a}
\begin{equation}
  \label{eq:chem-eq}
  \mu_0^l+\lambda^lC^l+k_BT\ln(C^l)=\mu_0^s+\lambda^sC^s+k_BT\ln(C^s),
\end{equation}
the $\mu_0$ and $\lambda$ parameters being constant and $k_B$ the
Boltzmann constant. For a given temperature, this equation allows one
to compute the concentration in the solid from the concentration in
the liquid. Note first that the equation is transcendental and must be
solved numerically, using Newton's method here. Second, the
concentration in the liquid is itself unknown and evolves according to
a conservation equation (eq.~\ref{eq:dxidt} in its mass fraction
expression) which itself depends on the partition coefficient that is
solution of equation~\eqref{eq:chem-eq}. In the case of O, as
discussed above, the evolution of the concentration in the liquid can
be predicted with great accuracy by assuming it is perfectly
incompatible. Solving equation~\eqref{eq:chem-eq} for each
concentration encountered in the liquid and the corresponding
temperature allows to compute the concentration in the solid.
\citet{Gubbins_etal2013} also decoupled the problem for S by assuming
a constant value for the concentration in the solid, which allows to
solve analytically the solute conservation equation and get the
evolution of the concentration in the liquid. Then they compute the
evolution of the concentration in the solid using the equilibrium
equation~\eqref{eq:chem-eq}.

In order to get closer to self-consistency, another approach is used
here. The thermal evolution of the core is modeled using the model
described in previous papers \citep{Labrosse03,Gomi_etal2013,Labrosse2014a} which
allows to compute the growth of the inner core with time. At each time
step, the new mass fraction of S in the liquid is obtained using
the conservation equation~\eqref{eq:dxidt} with the partition
coefficient obtained at the previous time step. The equilibrium
equation~\eqref{eq:chem-eq} is then used to get the concentration in
the solid newly accreted to the inner core. This also provides the new
value of the partition coefficient to be used in the next iteration. 

The equilibrium expressed by equation~\eqref{eq:chem-eq} only applies
to the ICB and not to the bulk of the outer and inner cores. Rejection
of light elements at the ICB drives convection in the outer core which
tends to stay well mixed, an assumption that was already made when
writing equation~\eqref{eq:xi-O-dev}. The other alternative proposed
by \citet{Alboussiere_etal2010} will be discussed later. For the inner
core, before convection sets in, the concentration can only be
homogenized by diffusion, a very slow process, particularly in the
solid and we ignore it altogether. \N{When the inner core is very
  small, diffusion can homogenize the solute, which would decrease the
  buoyancy available to drive convection. Neglecting diffusion
  therefore maximizes the chances for convection. With this
  approximation}, the change of solid concentration with time at the
ICB directly provides the profile as function of position in the inner
core.

The procedure explained above requires knowledge of the ICB temperature
for each inner core radius, which is equal to the liquidus of the
outer core composition, assumed uniform, and the corresponding
pressure. The liquidus is assumed to be only influenced by $\xi_O$,
not $\xi_S$, because of the vast difference in fractionation behavior \citep{Alfe_etal2007}.
Assuming that derivatives of the liquidus with pressure
($\partial T_L/\partial P$) and composition ($\partial
T_L/\partial\xi_O$) of
the liquidus are constant, the liquidus varies as function of the inner
core radius as 
\begin{equation}
  \label{eq:TL-r}
  T_L(r_{IC})=T_{L0}-K_0\left(\frac{\partial T_L}{\partial P}\right)_{\xi}\frac{r_{IC}^2}{L_\rho^2}
  +\left(\frac{\partial T_L}{\partial\xi_O}\right)_{P}\frac{\xi_{O0}^lr_{IC}^3}{r_{OC}^3\left(1-\frac{3r_{OC}^2}{5L_\rho^2}\right)},
\end{equation}
with $T_{L0}=T_L(0)$ the liquidus temperature at the central pressure for
the initial concentration of O in the core. \citet{Gubbins_etal2013}
did not consider the effect of composition on the liquidus. The value
of $T_{L0}$ is computed so that the liquidus for the present inner
core radius has the required value, as given in
table~\ref{tab:parameters} for the different compositional models
considered. \N{The effect of composition on the liquidus temperature,
  $\partial T_L/\partial\xi_O$, is taken from \citet{Alfe_etal2007}
  but converted to a mass fraction effect: for the PREM model, a liquidus deficit of
  650 K is due to a composition difference across the ICB in both O
  ($\Delta C_O=8\%$) and S ($\Delta C_S=2\%$) but O accounts for the
  largest part, 547K. The difference is mass fraction is 2.5\% and
  the ratio gives the value in table \ref{tab:parameters}. The effect of S on the
  liquidus is neglected here, as explained above.}

The evolution of the temperature follows from the diffusion equation
with a moving boundary at the ICB at which the liquidus temperature is imposed.
The moving boundary problem is solved using a front
fixing method \citep{Crank} by scaling the radius to that of the inner
core, $x=r/r_{IC}(t)$. We get an advection-diffusion equation,
\begin{equation}
  \label{eq:IC-diffusion}
  \rho(x) C_P\left(\frac{\partial T}{\partial
      t}-x\frac{\dot{r}_{IC}}{r_{IC}}\frac{\partial T}{\partial
      x}\right) = 
  \frac{1}{r_{IC}^2x^2}\frac{\partial}{\partial x}\left(x^2
  k(x) \frac{\partial T}{\partial x}\right),
\end{equation}
where $C_P$ is the heat capacity, the overdot means time derivative
and the thermal conductivity $k$ is assumed to vary spatially. This
equation is solved using a finite volume approach. The time step is
adapted so that the courant condition is satisfied at each time. The
boundary conditions are that of an imposed temperature at the ICB
($x=1$) and a zero flux at the center ($x=0$). 

The thermal conductivity is assumed to vary as a quadratic function of
radius \citep{Gomi_etal2013} which we write here as
\begin{equation}
  \label{eq:conductivity}
  k(r)=k_0\left(1-A_k\frac{r^2}{L_\rho^2}\right),
\end{equation}
with $k_0$ the central value and $A_k$ a constant. The link with the
more complex expression given in \citet{Gomi_etal2013} is
straightforward and the parameter values considered are given in
table~\ref{tab:parameters}.  \N{Note that the value considered here is on
the low end of the composition-dependent values proposed by
\citet{Gomi_etal2013} and correspond to the situation where Si is the
only light elements to explain the core density deficit and has the
same concentration in the inner core as in the outer core. This choice
is made in order to be conservative and maximize chances of convection
in the inner core. A more realistic value considering the effect of
the inner core composition would make the conductivity higher than
that considered here \citep{Pozzo_etal2014} and would render convection even less likely.}
\D{Note however that the conductivity of the
inner core might be larger than that of the outer core because of the
lower concentration of the inner core in impurities. This effect is
not considered to be on the conservative side. }

As appears clearly above, the knowledge of the growth history of the
inner core is sufficient to know the evolution of the ICB temperature
with time and therefore impose the required boundary condition for the
thermal diffusion and solve the chemical equilibrium required to
compute the composition profile. Therefore, a full model for the
thermal evolution of the outer core is not required. However, the
growth rate of the inner core is controlled by the CMB heat flow on
which some constraints exists, both from the mantle side
\citep{Jaupart_etal07} and from considerations on the stratification
of the core \citep{Gomi_etal2013,Pozzo_etal2012}. The model for the
inner core evolution is then implemented in the more general model for
the evolution of the outer core so that the whole evolution is driven
by an imposed history of CMB heat flow. The theory relating the inner
core growth to the CMB heat flow needs not be detailed here and can be
found elsewhere
\citep[eg][]{Braginsky_Roberts95,Labrosse03,Lister_Buffett95,Nimmo2007}. The
present paper uses a higher order version of the model presented in
\citet{Gomi_etal2013} that is discussed in another paper
\citep{Labrosse2014a}. It suffices here to state that the energy
balance of the core can be written as
\begin{equation}
  \label{eq:Energy-balance}
  Q_{CMB}=\Pc(r_{IC})\frac{dr_{IC}}{dt}+Q_{ICB},
\end{equation}
in which $\Pc$ is a function relating the radius of the inner core to
the sum of all energy sources of compositional, latent and sensible
origin. The heat flow across the ICB, $Q_{ICB}$, results from the
calculation of diffusion in the inner core, as presented above. Also,
\citet{Gomi_etal2013} showed that if the heat flow across the CMB,
$Q_{CMB}$, is lower than the isentropic value $Q_S$, a layer at the
top of the core would tend to become thermally stratified. The
thickness of the layer, larger than 1400km, is too important to go
unnoticed. The isentropic heat flow is therefore considered here
as a lower bound for the actual CMB heat flow. In practice, the heat
flow at the CMB is assumed to vary with time so that it keeps a
constant ratio to the isentropic value. 
\D{Other choices are
obviously possible but since the inner core age is found to be small, a
very large variation of the CMB heat flow is not expected over its
existence period. 
The different parameters used in the model are given in table~\ref{tab:parameters}.}

\section{Thermal stratification with a high \D{thermal} conductivity}
\label{sec:cond-therm-conv}
\begin{figure}[htbp]
  \centering
  \includegraphics{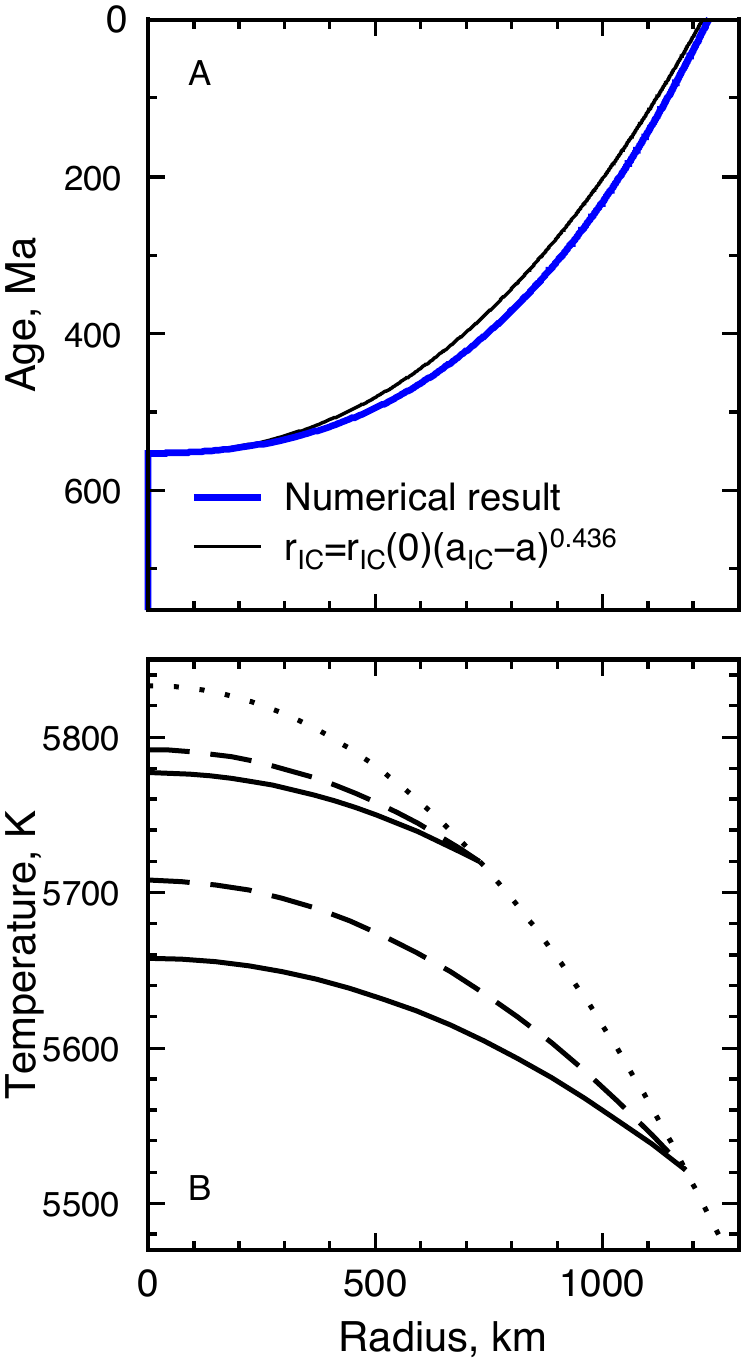}
  \caption{Example of inner core growth history (A) and temperature
    profiles (B) for a calculation assuming a CMB heat flow equal
    to 1.15 times the isentropic value, \N{that is $15.3\mathrm{TW}$ for
    the present time,} with the M\&G compositional
    model. The inner core growth is found to follow a power law with
    age $\tau$ with a power close to $1/2$ (thin black line). Profiles of conduction
    temperature in the inner core (solid lines on B) are generally
    found to be less steep than the isentropic ones (dashed). The
    liquidus profile (dotted) is computed as a function of inner core
    size and varies due to both pressure and composition.}
  \label{fig:TIC_final}
\end{figure}
Convection driven by secular cooling of the inner core has been
considered in great details by \citet{Deguen_Cardin2011} who derived
an approximated criterion for the temperature gradient to be
super-isentropic and therefore potentially unstable,
\begin{equation}
  \label{eq:stab-crit-deguen}
      \tau_{IC}<\frac{r_{IC}^2}{6\kappa}\left(\frac{dT_L}{dT_a}-1\right),
\end{equation}
with $\tau_{IC}$ the age of the inner core, $r_{IC}$ the present
radius of the inner core, $\kappa$ the thermal diffusivity and
$dT_L/dT_a$ the ratio of the liquidus to the isentropic
gradients. This criterion is based on the assumption of an inner core
growing as $\sqrt{t}$, which is a reasonable assumption. Indeed, as
shown on figure~\ref{fig:TIC_final}, the inner core growth is well
represented by a power law of time, with an exponent between $0.4$ and
$0.5$. The exact value depends on the composition since it affects the
evolution of the liquidus 

Using the parameters listed in table~\ref{tab:parameters},
\N{equation~\eqref{eq:stab-crit-deguen}} gives a maximum age for
thermal convection equal to 209 Myr. This age can be converted to a
CMB heat flow value of 40 TW. \N{Alternatively, since this criterion
  is approximate, the heat flow across the CMB can be increased so
  that the present temperature profile would make the inner core
  neutrally buoyant. Since the inner core always evolves toward
  stability even when starting unstably stratified
  \citep{Deguen_Cardin2011}, it would mean that the inner core would
  always have been unstably stratified. This happens for a CMB heat
  flow always equal to 2.2 times the isentropic value, 29 TW at
  present and an inner core age equal to 276 Myr. }  This value for
the CMB heat flow is unreasonably high considering the energy balance
of the mantle \citep{Jaupart_etal07}. For a more reasonable heat flow,
the temperature in the inner core is always sub-isentropic, as shown
on figure~\ref{fig:TIC_final}.

The recent upward revision of the thermal conductivity of the core
implies that the CMB heat flow must be larger than previously thought
for the dynamo to be convectively driven, at least using the
conventional buoyancy sources. This favors the possibility of inner
core convection since it implies a smaller inner core age than
previously envisioned \citep{Gomi_etal2013}. However, because the
thermal conductivity increases with depth in the core and with the
lesser amount of impurities in the inner core than the outer core, the
minimum requirements for thermal convection in the inner core are much
higher than the minimum requirements for convection in the outer core,
even without considering the effect of the vast difference in
viscosity. 
The strong stability implied by the type of temperature profile shown
on figure~\ref{fig:TIC_final} needs to be overcome by a compositional
instability for convection to occur in the inner core.





\section{Compositional stratification}
\label{sec:comp-strat}
\begin{figure}[htbp]
  \centering
  \includegraphics{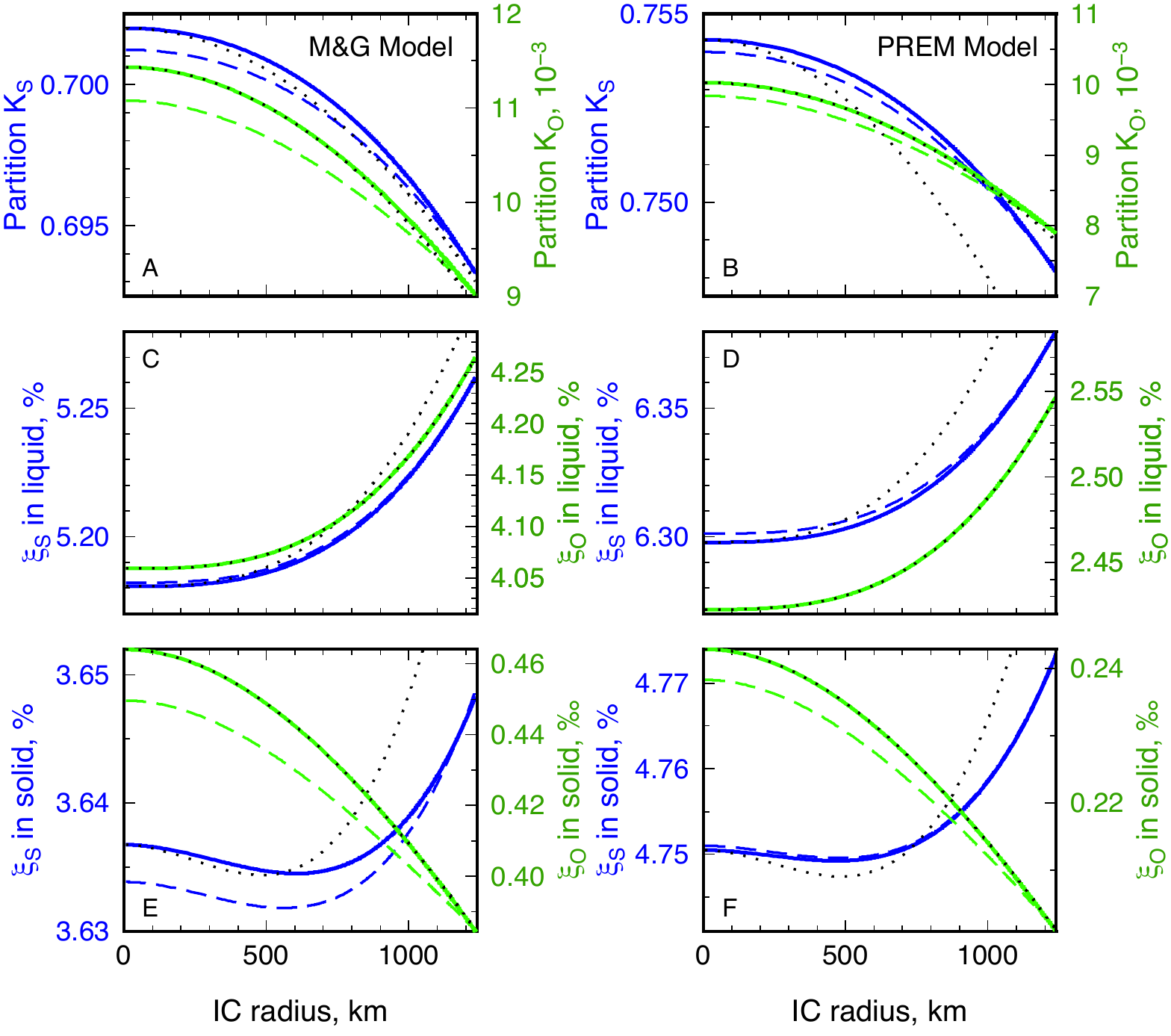}
    \caption{Evolution of the partition coefficient of S and O at the
      ICB (in the mass fraction sense, A, B), mass fraction of S and O
      in the liquid (C, D) and mass fraction of S and O in the solid
      (E, F) as function of the inner core radius for the two
      compositional models of \citet{Gubbins_etal2013}: M\&G model
      (A, C, E) and PREM model (B, D, F). Results for S are represented in
      blue with the axis on the left and for O in green with the axis
      on the right. Dotted lines are the variations predicted by
      leading order development which give \N{a} $r_{IC}^2$ \N{dependence} for
      the partition coefficients and \N{a} $r_{IC}^3$ \N{dependence} for the concentration
      in the liquid. \N{Dashed curves are obtained using the same
        parameterization as \citet{Gubbins_etal2013} for the
        concentration of S in the outer core and for the liquidus temperature.}}
  \label{fig:K-S}
\end{figure}
The decrease with time (inner core growth) of the ICB temperature from
both pressure and composition evolution leads to a change of partition
coefficient \citep{Gubbins_etal2013}. Figure~\ref{fig:K-S} shows the
decrease of the partition coefficient $K^{sl}$ for S and O with inner
core growth as well as the change of mass fraction of both elements in
the liquid and the solid, for the two compositional model of the Earth
core proposed by \citet{Gubbins_etal2013}. The general trends are
\N{qualitatively} similar for both models and the results differ only quantitatively. On
the other hand, the behavior is different between S and O: While the
mass fraction of O decreases with radius in the inner core, making it
prone to destabilization, the mass fraction of S starts by
decreasing before increasing. It appears that in the case of O the
effect of the change of partition coefficient is dominant for the
range of inner core size relevant to Earth while the effect of
increasing amount of S in the outer core dominates at large inner core
sizes. Overall, the stratification in $\xi_S$ will tend to make the
inner core stable while that in $\xi_O$ is adverse. 

\begin{table}[htbp]
  \centering
  \begin{tabular}{lllll}
    \hline
    Coefficient & M\&G model & PREM model & M\&G fit & PREM fit\\ 
    \hline 
    $A_O\ 10^{-11}\mathrm{km^{-3}}$ & $2.696$ & $2.696$ & $2.696$ &
    $2.696$ \\
    $A_S\ 10^{-11}\mathrm{km^{-3}}$ & $1.158$ & $0.884$ & $0.835$ & $0.687$\\
    $B_O\ 10^{-8}\mathrm{km^{-2}}$ & $14.51$& $14.66$ & $13.88$ & $13.85$
    \\
    $B_S\ 10^{-8}\mathrm{km^{-2}}$ &$0.837$ & $0.696$& $0.811$ & $0.532$\\
    $r_{IC,O}^\star\ \mathrm{km}$ &$3588$ & $3625$ & $3433$ & $3426$\\
    $r_{IC,S}^\star\ \mathrm{km}$ & $482$ & $525$ & $647$ & $517$\\
    \hline 
  \end{tabular}
  \caption{Coefficients of $r_{IC}$ in the leading order theory for
    the evolution of compositions and partition coefficients. The
    first two columns are the leading order terms and give the dotted
    lines on figure~\ref{fig:K-S} while the two last ones are fits of
    the full calculation using the same dependence on $r_{IC}$.}
  \label{tab:LowOrderTheory}
\end{table}
In order to understand this difference in behaviors, it is useful to
compute the leading order variations of the concentration and the
partition coefficient. In the case of O, equation~\eqref{eq:xi-O-dev}
provides the necessary expression, which gives
\begin{equation}
  \label{eq:DxiO-xiO}
  \frac{\delta\xi_O^l}{\xi_{O0}^l}=\frac{r_{IC}^3}{r_{OC}^3\left(1-\frac{3r_{OC}^2}{5L_\rho^2}\right)}
  \equiv A_O r_{IC}^3.
\end{equation}
In the case of S, assuming that, to leading order, the difference
of composition across the ICB $\Delta\xi_S^{ICB}$ is constant, one
gets
\begin{equation}
  \label{eq:DxiS-xiS}
  \frac{\delta\xi_S^l}{\xi_{S0}^l}=\frac{\Delta\xi_S^{ICB}}{\xi_{S0}}\frac{r_{IC}^3}{r_{OC}^3\left(1-\frac{3r_{OC}^2}{5L_\rho^2}\right)}
  \equiv A_S r_{IC}^3.
\end{equation}

For the partition coefficient, assuming negligible effect of the
variation of concentrations gives, from equation~\eqref{eq:chem-eq},
\begin{equation}
  \label{eq:PXsl}
  P_X^{sl}=\exp\left(\frac{\Delta\mu}{k_BT_L}\right)
  \simeq\exp\left(\frac{\Delta\mu}{k_BT_{L0}}\right)\left(1+\frac{\Delta\mu
    }{k_BT_{L0}}\frac{\delta T}{T_{L0}}\right),
\end{equation}
where $\Delta\mu=\mu_0^l-\mu_0^s+\lambda_X^lC_{X0}^l
-\lambda_X^sC_{X0}^s$ is the initial difference in chemical potential. 
The change of temperature is, from equation~\eqref{eq:TL-r}, dominated
by the pressure term which gives
\begin{equation}
  \label{eq:dP-P}
  \frac{\delta
    P_X^{sl}}{P_X^{sl}}=\frac{\Delta\mu}{k_BT_{L0}}\frac{K_0}{T_{L0}}\left(\frac{\partial
      T_L}{\partial P}\right)_{\xi}\frac{r_{IC}^2}{L_\rho^2}
  \equiv -B_X r_{IC}^2,
\end{equation}
where the minus sign has been introduced in order to make the
coefficient $B_X$ positive, since $\Delta\mu<0$.

Assuming that the logarithmic change of $K_X^{sl}$ is equal to that of
$P_X^{sl}$, we get for the change of concentration in the solid as
function of inner core radius:
\begin{equation}
  \label{eq:dxis}
  \frac{\delta\xi_X^s}{\xi_{X0}^s}=-B_X r_{IC}^2 + A_X r_{IC}^3.
\end{equation}
This equation shows that the decrease should dominate at small
$r_{IC}$ while the concentration should increase at large
$r_{IC}$. The switch of the two effects
happens for $r_{IC}=2B_X/3A_X\equiv r_{IC}^\star$ and the values of
the different coefficients and $r_{IC}^\star$ are listed in
table~\ref{tab:LowOrderTheory} for both compositional models. The
predictions from this leading order development are represented by
dotted lines on figure~\ref{fig:K-S} and show a very good agreement
with the full model in the case of O and a qualitatively good
agreement in the case of S. The value of $r_{IC}^\star$ match\N{es}
relatively well the full numerical result in the case of S but cannot
be tested in the case of O since it is larger than the core
radius. Alternatively, table~\ref{tab:LowOrderTheory} provides values
of the coefficients fitted to the full numerical results. 

Several implications need to be drawn from the results presented on
figure~\ref{fig:K-S}. First, O and S have very different behaviors. In
the case of O, the large value of $\Delta\mu_0$
(table~\ref{tab:parameters}) makes the partition coefficient vary more
strongly with temperature than the partition coefficient for S, by 1.5
orders of magnitudes. On the other hand, since the partition
coefficient is much smaller for O than for S, the increase of
the concentration in the liquid is larger for O than for S, but only
by a factor of 3 to 4. Therefore, while both effects balance for an
inner core about half its present size in the case of S, the effect of
the decrease in partition coefficient dominates up to the inner core
present size in the case of O. Even though O is much less present in
the inner core, it is more likely to make the inner core convect. This
contrasted behavior is found for both compositional models which only
differ quantitatively. 

It is difficult to compare with precision the present results to that
of \citet{Gubbins_etal2013} since they do not provide profiles of
concentrations in the same way as here and show their results as the
evolution of the Rayleigh number with inner core radius. Nevertheless,
their result is qualitatively similar to the ones presented here in
the case of the PREM model, with the concentration in S starting as
destabilizing and then stabilizing while the concentration in O is
always destabilizing. On the other hand, they find that both O and S
are always destabilizing in the case of the M\&G model, opposite to
what is shown on figure~\ref{fig:K-S}. The reason for this difference
in behavior is not clear. \N{A calculation including the approximate
  evolution of $C_S^l$ assumed by \citet{Gubbins_etal2013} and
  neglecting the effect of composition on the evolution of the
  liquidus temperature was performed and the results are shown as
  dashed lines on figure~\ref{fig:K-S}. The results are rather similar
  to the results obtained with the full model and the small difference
  comes mostly from the evolution of the liquidus temperature. In
  particular, the approximate solution of \citet{Gubbins_etal2013} for
  the evolution of the concentration in S the liquid appears to give rather
 good. This means that the qualitative differences between
  the present results and those from \citet{Gubbins_etal2013} in the
  case of the M\&G model cannot be explained by the differences in
  treatment of conservation of solute. Note however that the leading
  order analytical calculation presented above are qualitatively
  similar to the results of the full model, for both compositional
  models, and gives them support.}

Note that, since no compositional diffusion is considered in the inner
core (which maximizes the chances of convection), the composition
profile depends only on the inner core radius and not on any detail of
its growth rate. As explained in section~\ref{sec:cond-therm-conv},
the situation is different for the thermal stratification and the
combined thermal-compositional buoyancy requires to consider the time
evolution of the core.




\section{Combined buoyancy profiles and conditions for a unstable stratification}
\label{sec:comb-buoy-prof}
\begin{figure}[htbp]
  \centering
  \includegraphics{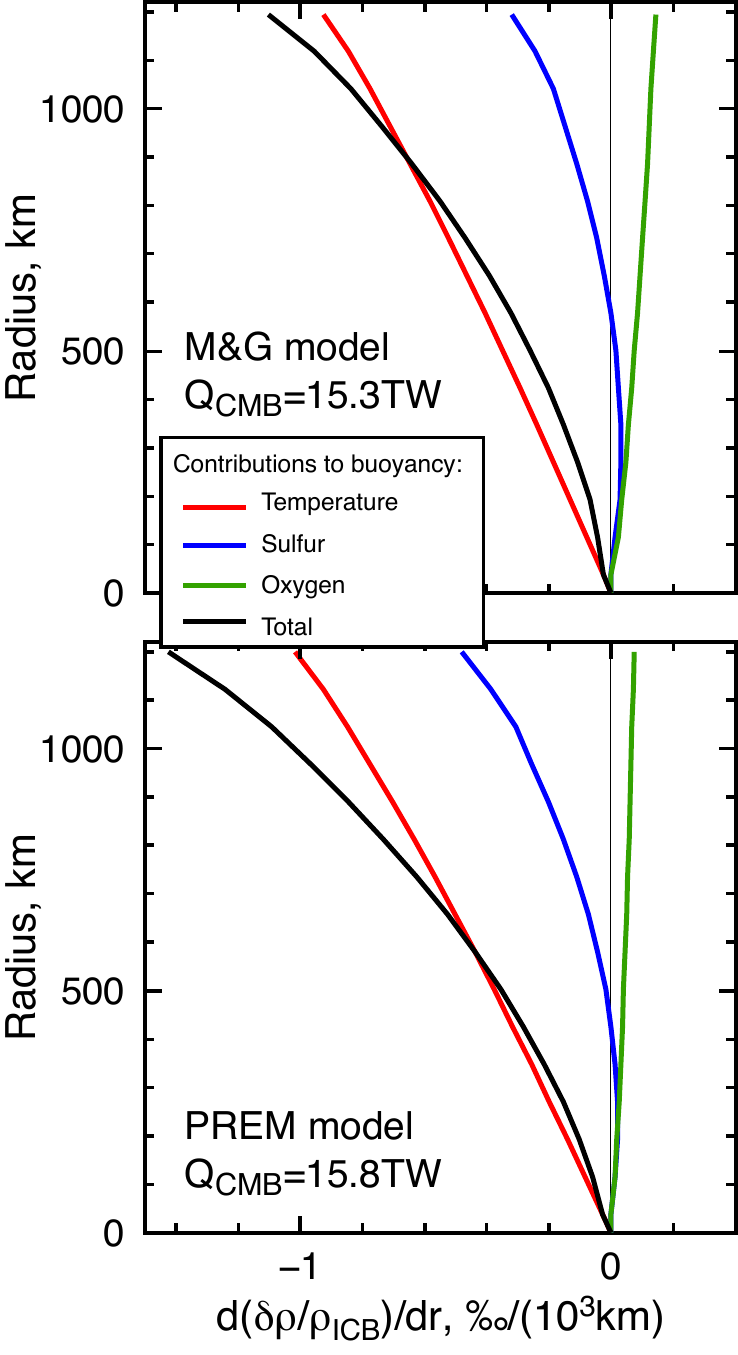}
  \caption{Vertical gradient of the buoyancy and its different
    contributions for the two compositionnal models. The CMB heat flow
    evolution is the same as that used to get
    figure~\ref{fig:TIC_final}\N{, always equal to 1.15 times the
      isentropic value and 15.3 TW and 15.8 TW at present in the M\&G
      and PREM models, respectively}.}
  \label{fig:buoyancy}
\end{figure}
For a reasonable CMB heat flow, the inner core has been found to be
stably stratified in the thermal sense. For composition, the situation
is less clear, with the concentration in O unstably stratified while
the concentration in S is potentially unstable in the innermost part
but stable in the outer part. The combined effect of both compositions
and temperature can be estimated by computing the density anomaly with
respect to the value at the ICB:
\begin{equation}
  \label{eq:buoyancy}
  \frac{\delta\rho(r)}{\rho_{ICB}}=-\alpha\left[T(r)-T_a(r)\right] 
  +\beta_S\left[\xi_S(r)-\xi_{Sf}\right] +\beta_O\left[\xi_O(r)-\xi_{Of}\right],
\end{equation}
with $T_a(r)$ the isentropic temperature profile, $\xi_{Xf}$ the mass
fraction of $X$ at the top of the inner core, $\beta_X$ the
corresponding expansion coefficient and $\alpha$ the coefficient of
thermal expansion (see table~\ref{tab:parameters} for parameter
values). The stratification is potentially unstable if the gradient
$\partial \delta\rho/\partial r>0$. Whether or not instability indeed
occurs in the case where this condition is fulfilled depends on
ill-constrained properties (diffusivities, viscosity) that all enter
in the Rayleigh number but it is first necessary to consider the sign of
the stratification.

Figure~\ref{fig:buoyancy} shows the different contributions,
temperature and concentration in O and S, to the vertical gradient of
the buoyancy, as defined in equation~\eqref{eq:buoyancy}. The thermal
part depends on the growth rate of the inner core, and the case
presented here corresponds to the calculation presented in
section~\ref{sec:cond-therm-conv}. 

It appears that, even though the concentration of O is always
destabilizing, its contribution to buoyancy is smaller than that of
S. The coefficient of chemical expansion is larger (in absolute sense)
for O than for S (table~\ref{tab:parameters}) but the amount of O
in the inner core is much smaller and so is its variation as
function of radius. For this reason, the destabilizing effect of O
cannot overcome the stabilizing effect of S. The resulting
compositional buoyancy is stabilizing in the case of the PREM model
and nearly neutral in the case of the M\&G model. The larger ICB
density jump proposed in the latter model than in PREM requires a
larger amount of O in the core and maximizes the importance of \N{the}
destabilizing \D{effect of} oxygen compared to the stabilizing \D{one of}
sulfur. Note however that this density jump is on the high end of all
proposals for this poorly constrained parameter
\citep{Hirose_etal2013}. Middle of the road values are closer to the
PREM number or even lower and would suggest a larger effect of S. In
any case, it appears that compositional buoyancy is not a very good
candidate to set the inner core in motion, at least within the
standard outer core model. Since thermal buoyancy is strongly
stabilizing, inner core convection seems hard to
sustain. Modifications of the standard scenario discussed above need
to be considered, however, to completely rule it out. Some
possibilities are mentioned in the next section.

\section{Discussion and conclusions}
\label{sec:disc-other-poss}
\N{The results presented above show that for any reasonable CMB heat flow
the thermal stratification is strongly stabilizing and that the
compositional stratification is at best neutral, at least when
considering an equilibrium between the inner core side of the ICB and
the bulk of the outer core, an alternate view being presented
below. But first, it is worth emphasizing that the choices of
thermal conduction parameters have been pushed systematically downward
in order to give convection the maximum chances. The values of the
conductivity parameters listed in table~\ref{tab:parameters} correspond
to Si being the only light element in the core
\citep{Gomi_etal2013}. Using a combination of S and O, as done for
other aspects of this paper, would make the central value of
conductivity $k_0\sim 215\mathrm{W/m/K}$ \citep{Gomi_etal2013}, for
the concentration assumed in the liquid. An even larger value should
be expected in the inner core since it contains less
solute. \citet{Pozzo_etal2014} give a conductivity at the center
equal to $237\mathrm{W/K/m}$ in that case. With this
value, a CMB heat flow 3.7 times larger than the isentropic (49 TW at
present) value is necessary to make the inner core unstably stratified
everywhere, corresponding to an inner core age equal to 185Myr. Such a
high CMB heat flow is clearly excluded and convection in the inner
core with such high thermal conductivity is impossible with a well
mixed outer core.}

\begin{figure}
  \centering
  \includegraphics{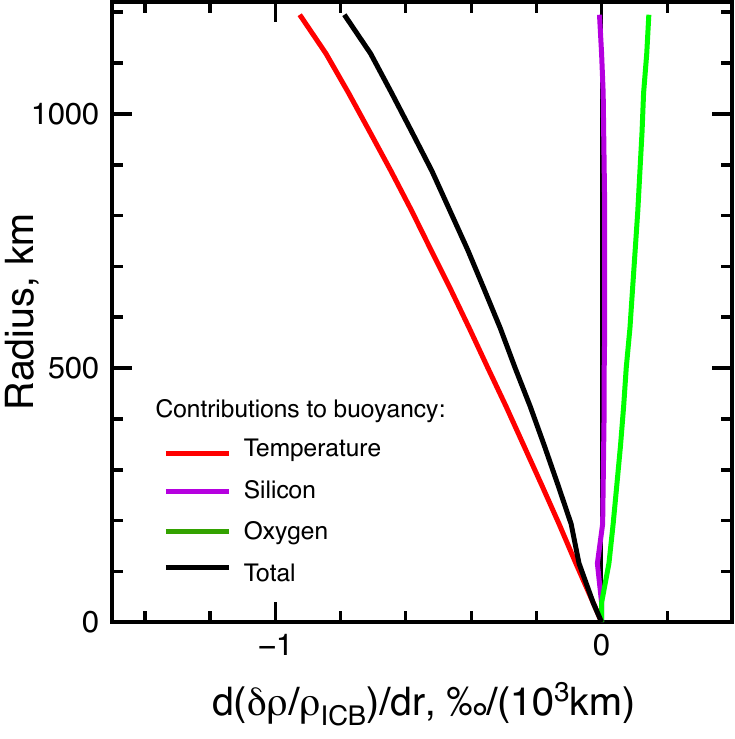}
  \caption{Final buoyancy in the case of a Fe-O-Si core composition
    with a partition coefficient equal to 1 for Si. The CMB heat flow
    is assumed to be equal to 1.15 times the isentropic value and
    equal to 15.3 TW and present.}
  \label{fig:buoyancy-Si}
\end{figure}
\N{The composition of core is still largely unknown and only two
  models have been considered above. These models were chosen because
  all the parameters needed to compute the evolution have been
  provided by previous studies \citep{Gubbins_etal2013}. It is quite
  possible that other choices of composition could change the results,
  although probably not enough to change completely the
  outcome. Another element, Si, is commonly considered as a
  possibility in the core. \citet{Alfe_etal02a} have computed
  equilibrium parameters and found no discernable partitioning at the
  ICB. This means that Si is not a good candidate to provide the
  buoyancy needed for inner core convection. On the other hand,
  since S is found to be stabilizing, if Si is considered in place of
  S in the core, as is an option to explain the density both for the
  inner core and the outer core \citep{Alfe_etal02a}, it could result
  in a more unstable situation. In order to estimate this effect, I
  ran some calculations where Si replaces S, keeping the same
  concentration but using the proper parameters and, in particular,
  a partition coefficient always equal to 1 and $\beta_{Si}=-0.91$ \citep{Deguen_Cardin2011}. 
  Since the M\&G model is the most likely to promote instability, I
  use these concentrations and just replace S by Si. The resulting
  buoyancy distribution is shown on figure~\ref{fig:buoyancy-Si}. Note
  that although the molar fraction of Si changes neither in the outer
  core nor in the inner core, since the partition coefficient is equal
to 1 and no expulsion results from the inner core growth, the
evolution of the concentration of O makes the mass fraction in Si
evolve (appendix~\ref{sec:mass-molar-fraction}). However, the
resulting buoyancy is negligible. The resulting total buoyancy is
strongly stabilizing for a reasonable CMB heat flow.}

The analysis presented above is based on the assumption that the outer
core is compositionally well mixed, which forms the basis of all
classical models of core dynamics and evolution. However, there
are some seismological evidences in favor of a compositional
stratification both at the base of the outer core \citep[the now
called F-layer, eg][]{Song_Helmberger1995,Souriau_Poupinet1991} and at
its summit
\citep[eg][]{Helffrich2010,Tanaka2007}. \citet{Alboussiere_etal2010}
proposed to explain the F-layer by a laterally varying
melting/freezing boundary condition at the ICB, owing to the
translation of the inner core. In this case, the equilibrium condition
represented by equation~\eqref{eq:chem-eq} applies to the liquid
adjacent to the inner core, not the bulk of the outer core and both
the model of \citet{Alboussiere_etal2010} and the condition of
stability of the F-layer argue for a liquid concentration at the ICB
lower than that of the bulk. If the formation of the F-layer results
from the crystallization of the inner core, the increasing
concentration in S and O of the bulk of the outer core would not
affect the concentration in the crystallizing solid and its evolution
would be dominated by the decrease of the partition
coefficient. However, if the F-layer formation mechanism requires a
convective instability, it is not clear how this process can ever
start. 

\begin{figure}[htbp]
  \centering
  \includegraphics{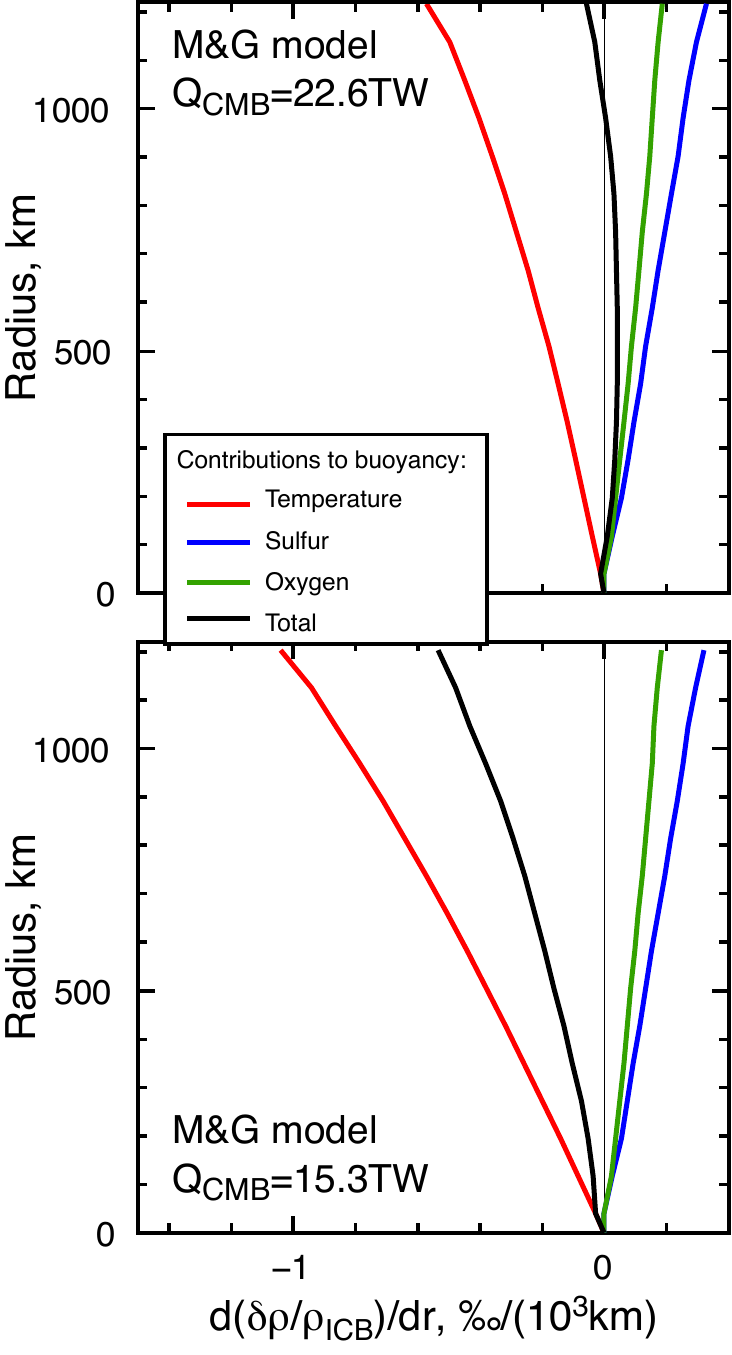}
  \caption{Buoyancy distribution assuming the concentration of the
    outer core does not vary with inner core growth (see text for
    details). The bottom panel is obtained for a CMB heat flow 1.15 times
    larger than the isentropic value (15.3TW at present) while the top
    panel is obtained for a CMB heat flow 1.7 times larger than the
    isentropic value (23.6TW at present).}
  \label{fig:F-layer}
\end{figure}
Alternatively, the evolution of the solute concentration of the outer
core can be affected by several processes occurring at the top of the
core. For example, barodiffusion can act to concentrate light elements
in a stably stratified layer \N{at the top of the} \D{in the uppermost} core
\citep{Fearn_Loper81b,Gubbins_Davies2013}. In this case, the
concentration of the bulk outer core in light elements is decreased
compared to the case where light elements are assumed to stay well
mixed. \citet{Buffett_etal00} also proposed that some of the light
elements contained in the core could sediment on its top, which would
lead to the same effect.  The outcome of the competition between the
solute concentration increase due to inner core growth and the
decrease from barodiffusion or sedimentation is not settled but
variations of the concentration in solute of the liquid just adjacent
to the inner core may not follow from the simple mass balance
considered in previous sections.

In order to get an idea of the importance of such effects, the
concentration in the liquid can be assumed to be constant with time,
meaning that either all the solute-rich fluid released at the ICB
by inner core growth is transported across the F-layer without
changing its composition or the flux of solute at the top of
the core toward the mantle or a stably stratified layer exactly
balances the flux from the inner core growth. 

Figure~\ref{fig:F-layer} shows the results of such calculations in
terms of the different contributions to the buoyancy distribution. The
bottom panel uses the same CMB heat flow history than the calculations
performed above, $Q_{CMB}=1.15Q_S$. Compar\N{ed} to the cases presented
in section~\ref{sec:comb-buoy-prof}, both O and S provide a
destabilizing buoyancy, but this is still not sufficient to make the
total density structure unstably stratified. In order to get neutrally
buoyant, the CMB heat flow must be \N{1.7} times the isentropic value,
that is \N{22.6} TW at present, as can be seen on the top panel of
figure~\ref{fig:F-layer}. This value \N{would imply that the mantle
is essentially not cooling, which contradicts observations from basalt
chemistry} \citep{Jaupart_etal07}. 

Note that if the concentration in solute at the bottom of the outer
core is kept constant with time, the decrease of the liquidus
temperature with time is lessened compared to the case where the outer
core is assumed well mixed. This decreases the effect of temperature
on the partition coefficient and this explains why density
stratification is not made more dramatically unstable. 

For compositional convection to occur in the inner core, it appears
that the concentration of solutes at the bottom of the outer core must
decrease with time. The mechanism proposed by
\citet{Alboussiere_etal2010} may allow that but requires inner core
convection, possibly in the form of translation. The density
stratification computed in various cases here appears stable for
reasonable values of the CMB heat flow. However, because of the vast
difference between thermal and chemical diffusivities,
double-diffusive instabilities might still be possible
\citep{Turner73}. \N{However, as pointed out by \citet{Pozzo_etal2014},
the timescale for the growth of this instability is the thermal diffusion
one and is similar to the age of the inner core.} This option might
still be the last remaining chance for convection in the inner core
and needs to be tested in the future.

\section{Acknowledgments}
\label{sec:acknowledgments}
Suggestions from Renaud Deguen, discussions with Marine Lasbleis,
Thierry Alboussi\`ere, Yanick Ricard, Dario Alf\`e and reviews by Chris Davies and
Bruce Buffett were very helpful in preparing this paper.

\appendix
\section{Mass and molar fraction}
\label{sec:mass-molar-fraction}
Depending on the context, molar or mass fraction of any light element is used.
Specifically, molar fractions are used by \citet{Gubbins_etal2013} for their model of
chemical equilibrium at the ICB but mass fraction are more convenient for the thermal
evolution model \cite[eg][]{Braginsky_Roberts95,Labrosse03}. Expressions to go from one system to the other are
provided here.

Let $x_i$ denote the mole fraction of element $i$ (molar mass $M_i$) in the mixture, its
molar concentration is $C_i=x_i\rho/M$, with $\rho$ the density of the mixture and $M=\sum_ix_iM_i$ the
average molar mass. The mass fraction of element $i$ is $\xi_i=x_iM_i/M$. For a mixture of
$N$ species only $N-1$ independent fractions define the composition. Considering, as \citet{Gubbins_etal2013}, a
tertiary mixture of Fe, S and O, the mass fractions in S and O simply write as
\begin{align}
  \label{eq:xi-O-x}
  \xi_O&=\frac{x_OM_O}{x_OM_O+x_SM_S+(1-x_O-x_S)M_{Fe}},\\
  \label{eq:xi-S-x}
  \xi_S&=\frac{x_SM_S}{x_OM_O+x_SM_S+(1-x_O-x_S)M_{Fe}}.
\end{align}
Inversion of this set of equations leads expressions of $x_i$ as functions of $\xi_i$:
\begin{align}
  \label{eq:x-O}
  x_O&=\frac{\xi_OM_SM_{Fe}}{(\xi_OM_S+\xi_SM_O)M_{Fe}+(1-\xi_O-\xi_S)M_OM_S},\\
  \label{eq:x-S}
  x_S&=\frac{\xi_SM_OM_{Fe}}{(\xi_OM_S+\xi_SM_O)M_{Fe}+(1-\xi_O-\xi_S)M_OM_S}.
\end{align}

\bibliographystyle{cras2a}
\bibliography{jrn,biblio,bouquin}

\begin{thebibliography}{42}
\providecommand{\natexlab}[1]{#1}
\expandafter\ifx\csname urlstyle\endcsname\relax
  \providecommand{\doi}[1]{doi:\discretionary{}{}{}#1}\else
  \providecommand{\doi}{doi:\discretionary{}{}{}\begingroup
  \urlstyle{rm}\Url}\fi

\bibitem[{{Alboussi\`ere} et~al.(2010){Alboussi\`ere}, Deguen and
  Melzani}]{Alboussiere_etal2010}
{Alboussi\`ere}, T., Deguen, R., Melzani, M., 2010.
\newblock Melting-induced stratification above the {Earth}'s inner core due to
  convective translation.
\newblock Nature, 466, 744--747.

\bibitem[{{Alf\`e} et~al.(1999){Alf\`e}, Gillan and Price}]{Alfe_etal99}
{Alf\`e}, D., Gillan, M.J., Price, G.D., 1999.
\newblock The melting curve of iron at the pressures of the {Earth}'s core from
  ab initio calculations.
\newblock Nature, 401, 462--463.

\bibitem[{{Alf\`e} et~al.(2002){Alf\`e}, Gillan and Price}]{Alfe_etal02a}
{Alf\`e}, D., Gillan, M.J., Price, G.D., 2002.
\newblock Composition and temperature of the {Earth's} core constrained by
  combining ab initio calculations and seismic data.
\newblock Earth Planet. Sci. Lett., 195, 91--98.

\bibitem[{Alf{\`e} et~al.(2007)Alf{\`e}, Gillan and Price}]{Alfe_etal2007}
Alf{\`e}, D., Gillan, M.J., Price, G.D., 2007.
\newblock Temperature and composition of the {Earth}'s core.
\newblock Contemporary Physics, 48, 63--80.
\newblock \doi{10.1080/00107510701529653}.

\bibitem[{Braginsky and Roberts(1995)}]{Braginsky_Roberts95}
Braginsky, S.I., Roberts, P.H., 1995.
\newblock Equations governing convection in {Earth}'s core and the geodynamo.
\newblock Geophys. Astrophys. Fluid Dyn, 79, 1--97.

\bibitem[{Buffett(2009)}]{Buffett2009}
Buffett, B.A., 2009.
\newblock Onset and orientation of convection in the inner core.
\newblock Geophys. J. Int., 179, 711--719.
\newblock \doi{10.1111/j.1365-246X.2009.04311.x}.

\bibitem[{Buffett et~al.(2000)Buffett, Garnero and Jeanloz}]{Buffett_etal00}
Buffett, B.A., Garnero, E.J., Jeanloz, R., 2000.
\newblock Sediments at the top of {Earth}'s core.
\newblock Science, 290, 1338--1342.

\bibitem[{Cottaar and Buffett(2012)}]{Cottaar_Buffett2012}
Cottaar, S., Buffett, B., 2012.
\newblock Convection in the {Earth}'s inner core.
\newblock Phys. Earth Planet. Inter., 198--199, 67--78.

\bibitem[{Crank(1984)}]{Crank}
Crank, J., 1984.
\newblock Free and moving boundary problems.
\newblock Oxford University Press, Oxford.
\newblock 425 pp.

\bibitem[{de~Koker et~al.(2012)de~Koker, Steinle-Neumann and Vl{\v
  c}ek}]{deKoker_etal2012}
de~Koker, N., Steinle-Neumann, G., Vl{\v c}ek, V., 2012.
\newblock Electrical resistivity and thermal conductivity of liquid {Fe} alloys
  at high {P} and {T}, and heat flux in {Earth}'s core.
\newblock Proc. Nat. Acad. Sci. U.S.A., 109, 4070--4073.

\bibitem[{Deguen(2012)}]{Deguen2012}
Deguen, R., 2012.
\newblock Structure and dynamics of {Earth}'s inner core.
\newblock Earth Planet. Sci. Lett., 333--334, 211 -- 225.
\newblock \doi{10.1016/j.epsl.2012.04.038}.

\bibitem[{Deguen et~al.(2013)Deguen, Alboussi{\`e}re and
  Cardin}]{Deguen_etal2013}
Deguen, R., Alboussi{\`e}re, T., Cardin, P., 2013.
\newblock Thermal convection in {Earth}'s inner core with phase change at its
  boundary.
\newblock Geophys. J. Int., 194, 1310--1334.
\newblock \doi{10.1093/gji/ggt202}.

\bibitem[{Deguen and Cardin(2011)}]{Deguen_Cardin2011}
Deguen, R., Cardin, P., 2011.
\newblock Thermochemical convection in {Earth}'s inner core.
\newblock Geophys. J. Int., 187, 1101--1118.
\newblock \doi{10.1111/j.1365-246X.2011.05222.x}.

\bibitem[{Dziewonski and Anderson(1981)}]{PREM}
Dziewonski, A.M., Anderson, D.L., 1981.
\newblock Preliminary reference {Earth} model.
\newblock Phys. Earth Planet. Inter., 25, 297--356.

\bibitem[{Fearn and Loper(1981)}]{Fearn_Loper81b}
Fearn, D.R., Loper, D.E., 1981.
\newblock Compositional convection and stratification of {Earth}'s core.
\newblock Nature, 289, 393--394.

\bibitem[{Gomi et~al.(2013)Gomi, Ohta, Hirose, Labrosse, Caracas, Verstraete
  and Hernlund}]{Gomi_etal2013}
Gomi, H., Ohta, K., Hirose, K., Labrosse, S., Caracas, R., Verstraete, M.J.,
  Hernlund, J.W., 2013.
\newblock The high conductivity of iron and thermal evolution of the {Earth}'s
  core.
\newblock Phys. Earth Planet. Inter., 224, 88--103.
\newblock \doi{10.1016/j.pepi.2013.07.010}.

\bibitem[{Gubbins et~al.(2013)Gubbins, Alf{\`e} and Davies}]{Gubbins_etal2013}
Gubbins, D., Alf{\`e}, D., Davies, C.J., 2013.
\newblock Compositional instability of {Earth}'s solid inner core.
\newblock Geophys. Res. Lett., 40, 1084--1088.
\newblock \doi{10.1002/grl.50186}.

\bibitem[{Gubbins and Davies(2013)}]{Gubbins_Davies2013}
Gubbins, D., Davies, C., 2013.
\newblock The stratified layer at the core--mantle boundary caused by
  barodiffusion of oxygen, sulphur and silicon.
\newblock Phys. Earth Planet. Inter., 215, 21 -- 28.
\newblock \doi{10.1016/j.pepi.2012.11.001}.

\bibitem[{Gubbins et~al.(2003)Gubbins, {Alf\`e}, Masters, Price and
  Gillan}]{Gubbins_etal03}
Gubbins, D., {Alf\`e}, D., Masters, G., Price, G.D., Gillan, M.J., 2003.
\newblock Can the {Earth}'s dynamo run on heat alone?
\newblock Geophys. J. Int., 155, 609--622.

\bibitem[{Helffrich and Kaneshima(2010)}]{Helffrich2010}
Helffrich, G., Kaneshima, S., 2010.
\newblock Outer-core compositional stratification from observed core wave speed
  profiles.
\newblock Nature, 468, 807--810.

\bibitem[{Hirose et~al.(2013)Hirose, Labrosse and Hernlund}]{Hirose_etal2013}
Hirose, K., Labrosse, S., Hernlund, J.W., 2013.
\newblock Composition and state of the core.
\newblock Ann. Rev. Earth Planet. Sci., 41, 657--691.
\newblock \doi{10.1146/annurev-earth-050212-124007}.

\bibitem[{Jaupart et~al.(2007)Jaupart, Labrosse and Mareschal}]{Jaupart_etal07}
Jaupart, C., Labrosse, S., Mareschal, J.C., 2007.
\newblock Treatise on Geophysics. Mantle dynamics., Elsevier, vol.~7, chap.
  6-Temperatures, Heat and Energy in the Mantle of the Earth.
\newblock pp. 253--303.

\bibitem[{Jeanloz and Wenk(1988)}]{Jeanloz_Wenk88}
Jeanloz, R., Wenk, H.R., 1988.
\newblock Convection and anisotropy of the inner core.
\newblock Geophys. Res. Lett., 15, 72--75.

\bibitem[{Labrosse(2003)}]{Labrosse03}
Labrosse, S., 2003.
\newblock Thermal and magnetic evolution of the {Earth}'s core.
\newblock Phys. Earth Planet. Inter., 140, 127--143.

\bibitem[{{Labrosse}(2014)}]{Labrosse2014a}
{Labrosse}, S., 2014.
\newblock Thermal evolution of the core with a high thermal conductivity.
\newblock Phys. Earth Planet. Inter.
\newblock In prep.

\bibitem[{Labrosse et~al.(2001)Labrosse, Poirier and
  Le~{Mou\"el}}]{Labrosse_etal01}
Labrosse, S., Poirier, J.P., Le~{Mou\"el}, J.L., 2001.
\newblock The age of the inner core.
\newblock Earth Planet. Sci. Lett., 190, 111--123.

\bibitem[{Lister and Buffett(1995)}]{Lister_Buffett95}
Lister, J.R., Buffett, B.A., 1995.
\newblock The strength and efficiency of the thermal and compositional
  convection in the geodynamo.
\newblock Phys. Earth Planet. Inter., 91, 17--30.

\bibitem[{Masters and Gubbins(2003)}]{Masters_Gubbins03}
Masters, G., Gubbins, D., 2003.
\newblock On the resolution of density within the {Earth}.
\newblock Phys. Earth Planet. Inter., 140, 159--167.

\bibitem[{Mizzon and Monnereau(2013)}]{Mizzon_Monnereau2013}
Mizzon, H., Monnereau, M., 2013.
\newblock Implication of the lopsided growth for the viscosity of {Earth}'s
  inner core.
\newblock Earth Planet. Sci. Lett., 361, 391 -- 401.
\newblock \doi{10.1016/j.epsl.2012.11.005}.

\bibitem[{Morelli et~al.(1986)Morelli, Dziewonski and
  Woodhouse}]{Morelli_etal1986}
Morelli, A., Dziewonski, A.M., Woodhouse, J.H., 1986.
\newblock Anisotropy of the inner core inferred from {PKIKP} travel times.
\newblock Geophys. Res. Lett., 13, 1545--1548.
\newblock \doi{10.1029/GL013i013p01545}.

\bibitem[{Nimmo(2007)}]{Nimmo2007}
Nimmo, F., 2007.
\newblock Treatise on Geophysics, Elsevier, vol.~8, chap. 2. Energetics of the
  core.
\newblock pp. 253--303.

\bibitem[{Poupinet et~al.(1983)Poupinet, Pillet and Souriau}]{Poupinet1983}
Poupinet, G., Pillet, R., Souriau, A., 1983.
\newblock Possible heterogeneity of the {Earth}'s core deduced from {PKIKP}
  travel times.
\newblock Nature, 305, 204--206.

\bibitem[{Pozzo et~al.(2014)Pozzo, Davies, Gubbins and
  Alf{\`e}}]{Pozzo_etal2014}
Pozzo, M., Davies, C., Gubbins, D., Alf{\`e}, D., 2014.
\newblock Thermal and electrical conductivity of solid iron and iron--silicon
  mixtures at earth's core conditions.
\newblock Earth and Planetary Science Letters, 393, 159 -- 164.
\newblock \doi{10.1016/j.epsl.2014.02.047}.

\bibitem[{Pozzo et~al.(2012)Pozzo, Davies, Gubbins and
  Alf{\`e}}]{Pozzo_etal2012}
Pozzo, M., Davies, C.J., Gubbins, D., Alf{\`e}, D., 2012.
\newblock {Thermal and electrical conductivity of iron at Earth's core
  conditions}.
\newblock Nature, 485, 355--358.

\bibitem[{Song and Helmberger(1995)}]{Song_Helmberger1995}
Song, X., Helmberger, D.V., 1995.
\newblock A {P} wave velocity model of {Earth}'s core.
\newblock J. Geophys. Res., 100, 9817--9830.
\newblock \doi{10.1029/94JB03135}.

\bibitem[{Souriau and Poupinet(1991)}]{Souriau_Poupinet1991}
Souriau, A., Poupinet, G., 1991.
\newblock The velocity profile at the base of the liquid core from
  {PKP(BC+Cdiff)} data: An argument in favour of radial inhomogeneity.
\newblock Geophys. Res. Lett., 18, 2023--2026.
\newblock \doi{10.1029/91GL02417}.

\bibitem[{Takehiro(2011)}]{Takehiro2011}
Takehiro, S.I., 2011.
\newblock Fluid motions induced by horizontally heterogeneous {Joule} heating
  in the {Earth}'s inner core.
\newblock Phys. Earth Planet. Inter., 184, 134 -- 142.
\newblock \doi{10.1016/j.pepi.2010.11.002}.

\bibitem[{Tanaka(2007)}]{Tanaka2007}
Tanaka, S., 2007.
\newblock {Possibility of a low P-wave velocity layer in the outermost core
  from global SmKS waveforms}.
\newblock Earth Planet. Sci. Lett., 259, 486--499.
\newblock \doi{10.1016/j.epsl.2007.05.007}.

\bibitem[{Turner(1973)}]{Turner73}
Turner, J.S., 1973.
\newblock Buoyancy Effects in Fluids.
\newblock Cambridge University Press, Cambridge.
\newblock 368 pp.

\bibitem[{Weber and Machetel(1992)}]{Weber_Machetel92}
Weber, P., Machetel, P., 1992.
\newblock Convection within the inner-core and thermal implications.
\newblock Geophys. Res. Lett., 19, 2107--2110.

\bibitem[{Woodhouse et~al.(1986)Woodhouse, Giardini and
  Li}]{Woodhouse_etal1986}
Woodhouse, J.H., Giardini, D., Li, X.D., 1986.
\newblock Evidence for inner core anisotropy from free oscillations.
\newblock Geophys. Res. Lett., 13, 1549--1552.
\newblock \doi{10.1029/GL013i013p01549}.

\bibitem[{Yukutake(1998)}]{Yukutake1998}
Yukutake, T., 1998.
\newblock Implausibility of thermal convection in the {Earth}'s solid inner
  core.
\newblock Phys. Earth Planet. Inter., 108, 1 -- 13.
\newblock \doi{10.1016/S0031-9201(98)00097-1}.

\end{thebibliography}
\end{document}